\documentclass[%
 aip,
 amsmath,amssymb,
 reprint,%
]{revtex4-1}

\usepackage{graphicx}
\usepackage{dcolumn}
\usepackage{bm}
\usepackage{color}
\usepackage[utf8]{inputenc}
\usepackage[T1]{fontenc}
\usepackage{mathptmx}
\usepackage{etoolbox}

\makeatletter
\def\@email#1#2{%
 \endgroup
 \patchcmd{\titleblock@produce}
  {\frontmatter@RRAPformat}
  {\frontmatter@RRAPformat{\produce@RRAP{*#1\href{mailto:#2}{#2}}}\frontmatter@RRAPformat}
  {}{}
}%
\makeatother
\begin{document}

\preprint{AIP/123-QED}

\title[Rheology of {\it Pseudomonas fluorescens} biofilms]{Rheology of {\it Pseudomonas fluorescens} biofilms: from experiments to predictive DPD mesoscopic modelling.}

\author{Jos\'e Mart\'in-Roca}
\affiliation{Departamento de Estructrura de la Materia, Física Térmica y Electrónica, Universidad Complutense de Madrid, 28040 Madrid, Spain}
 \affiliation{Departamento de Quimica Fisica, Universidad Complutense de Madrid, 28040 Madrid, Spain}

\author{Valentino Bianco}
\affiliation{Departamento de Estructrura de la Materia, Física Térmica y Electrónica, Universidad Complutense de Madrid, 28040 Madrid, Spain}
\affiliation{xxxx}

\author{Francisco Alarc\'on}
\affiliation{Departamento de Ingeniería Física, División de Ciencias e Ingenierías, Universidad de Guanajuato, Loma del Bosque 103, 37150 León, Mexico.}

\author{Ajay K. Monnappa}
\affiliation{Instituto de Investigación Biomédica Hospital Doce de Octubre (imas12), 28041 Madrid, Spain.}

\author{Paolo Natale}
\affiliation{Departamento de Quimica Fisica, Universidad Complutense de Madrid, 28040 Madrid, Spain}
\affiliation{
Instituto de Investigación Sanitaria Hospital Doce de Octubre (imas12), 28041 Madrid, Spain.}

\author{Francisco Monroy}
\affiliation{Translational Biophysics. Instituto de Investigación Sanitaria Hospital Doce de Octubre (Imas12), 28041 Madrid, Spain.}
\affiliation{Biophysics for Biotechnology and Biomedicine (Biophys-HUB). Departamento de Química Física, Universidad Complutense de Madrid, 28040 Madrid, Spain}

\author{Belen Orgaz}
\affiliation{Sección Departamental de Farmacia Galénica y Tecnología Alimentaria, Universidad Complutense de Madrid, Madrid, Spain}
\author{Ivan L\'opez-Montero}

\affiliation{Instituto de Investigación Sanitaria Hospital Doce de Octubre (imas12), 28041 Madrid, Spain.}
\affiliation{Departamento de Química Fisica, Universidad Complutense de Madrid, 28040 Madrid, Spain}

\author{Chantal Valeriani}
\affiliation{Departamento de Estrcutrura de la Materia, Física Térmica y Electrónica, Universidad Complutense de Madrid, 28040 Madrid, Spain}
\affiliation{Corresponding author}

\date{\today}

\begin{abstract}

{\color{black} Bacterial biofilms mechanically behave as viscoelastic media consisting of micron-sized bacteria crosslinked to a self-produced network of extracellular polymeric substances (EPS) embedded in water. Structural principles for numerical modelling aim at describing mesoscopic viscoelasticity without loosing detail on the underlying interactions existing in wide regimes of deformation under hydrodynamic stress. Here we approach the computational challenge to model bacterial biofilms for predictive mechanics {\it in silico} under variable stress conditions. Up-to-date models are not entirely satisfactory due to the plethora of parameters required to make them functioning under the effects of stress. As guided by the structural depiction gained in a previous work with {\it Pseudomonas fluorescens} (Jara et al. Front. Microbiol. (2021)), we propose a mechanical modeling by means of Dissipative Particle Dynamics (DPD), which captures the essentials of the topological and compositional interactions between bacteria particles and crosslinked EPS-embedding under imposed shear. The {\it P. fluorescens} biofilms have been modeled under mechanical stress mimicking shear stresses as undergone {\it in vitro}. The predictive capacity for mechanical features in DPD-simulated biofilms has been investigated by varying the externally imposed field of shear strain at variable amplitude and frequency. The parametric map of essential biofilm ingredients has been explored by making the rheological responses to emerge among conservative mesoscopic interactions and frictional dissipation in the underlying microscale. The proposed coarse grained DPD simulation qualitatively catches the rheology of the {\it P. fluorescens} biofilm over several decades of dynamic scaling.}  
\end{abstract}

\maketitle

\section{\label{sec:intro}Introduction}

Bacterial biofilms are synergistic colonies growing on solid surfaces in contact with a complex aqueous phase that provides physical protection against external threats\cite{peterson,fleming2010}. Because  biofilm colonies are mechanically more resilient than the isolated bacteria, they become more capable to nurse resistant phenotypes\cite{donlan2001biofilms,Hoiby2010}. A collective efficiency emerges from the microbial colony to engender antibiotic resistances and compromise substrate removal causing hence persistent biofilm infection\cite{donlan2001biofilms,Hoiby2010}. Surface biofilm formation is indeed a major health issue\cite{Khatoon2018, Milchev2013} and an industrial challenge worldwide\cite{Duan2008,Mansour2016,Schultz2000}. The mesoscopic structure of biofilms consists of a flexible meshwork made of bacteria and extracellular polymer substance (EPS), which is composed of polysaccharides, proteins and extracellular DNA embedded in aqueous suspension. Depending on the bacterial species, the amount of thriving cells with respect to the EPS vary from one tenth to one fourth of the total biofilm mass\cite{Sveinbjornsson2012}. {\color{black} Biofilms are considered to be complex materials as highly crosslinked EPS composites with a heterogeneous viscoelasticity acting as a dynamic scaffold for the dwelling cells \cite{wilking2011,jara2021}. Recently, biofilm viscoelasticity has been considered for probing the mechanical interlinking between EPS components and dwelling bacteria\cite{jara2021,boudarel2018towards,samuel2019regulating,jana2020nonlinear,billings2015material,gordon2017biofilms}. As a complexity emerging from compositional and topological configurations in the EPS assembly at multiple length scales\cite{boudarel2018towards, klauck2018spatial, stewart2015artificial}, biofilms exhibit both elasticity and fluid-like rheological response upon shear arising from their complex hydrodynamic dependence on the deformation amplitudes and frequencies\cite{samuel2019regulating, jana2020nonlinear, billings2015material, gordon2017biofilms}. Relevant structural interactions such as bacterial entanglement, protein binding, and EPS cross-linking contribute to form the transient stress-bearing structure that makes the particular biofilms' viscoelasticity to emerge from the mesoscale \cite{ganesan2013molar,vergara2009relevant,jennings2015pel}. Recent structural studies have shown that EPS constituents, such as secreted polysaccharides, DNA chains and other protein filaments biochemically dictate matrix architecture as well as biofilm stability under low mechanical load in the linear regime\cite{ganesan2013molar, vergara2009relevant, jennings2015pel}. However, there is a lack of understanding of the cross-linking, dynamical rearrangements, and mesoscopic reorganizations under large shear forces mimicking real-thing perturbations. Rheological techniques for probing large-amplitude shear deformation have allowed to record rheological signatures at a variety of stresses under wide configurational landscape\cite{jara2021,samuel2019regulating,billings2015material,gordon2017biofilms,sankaran2018quantitative}, and strong history fidelity (i.e., presenting viscoelastic memory) \cite{jana2020nonlinear}. The biofilms grown under different mechanical conditions are indeed known to exhibit viscoelastic variations as corresponding to environmental adaptations\cite{jara2021, tallawi2017modulation}. Furthermore, EPS viscoelasticity is known to confer protection against chemical and physical threats \cite{peterson2015viscoelasticity, gloag2018viscoelastic}. Whilst ample literature alludes to the biomolecular role of the EPS secreted under stressed biofilm growth \cite{jara2021,tallawi2017modulation,peterson2015viscoelasticity}, systematic investigations of the rheological response are still lacking on the biophysical focus. To investigate {\it in silico} a synthetic biofilms’ rheology that improves further our predictive ability from structural coarse-grain standpoints, we invoke numerical modelling from mesoscopic physics as approached alongside experimental rheology data obtained in physicochemically controlled biofilms \cite{jana2020nonlinear}.}\\

Physical biofilm modelling started more than one decade ago in terms of continuous media approaches\cite{klapper2010,Wang2010}. They were first described via phase-field (continuum) models in which the disperse phase accounts for polymerized biofilm components (EPS and bacteria), and the continuous phase for aqueous ambiance (containing nutrients and many other small molecules)\cite{zang2008phase}. This model has been successfully used to recapitulate essentials of biofilm growth, particularly unraveling mechanical features in agarose hydrogels containing bacteria \cite{kandemir}. A further approach consisted in implementing hybrid discrete-continuum models to couple bacterial growth and biomass spreading over biofilm roughness \cite{Picioreanu1999,Picioreanu2004}. Numerical mesoscale approaches have been also proposed, particularly Dissipative Particle Dynamics (DPD\cite{groot1997dissipative,espanol2017perspective}), and the Immersed Boundary-based Method (IBM)\cite{peskin1,peskin2}.  IBM relies on strong assumptions such as: a) considering biofilm steadiness during the simulation time scale \cite{IBM0}; b) imposing biofilm elasticity as linear springs connecting individual bacteria \cite{IBM}; c) capturing the biofilm viscosity via constitutive stresses. Encoding these constraints has allowed using IBM for studying biofilm growth and its force-guided deformation \cite{Horn2014,Eberl2008,Liu2018,Xu2011}, establishing connections between mechanical stress and biofilm strain \cite{Bol2012,Hammo2014,Stotsky2016}. The DPD methods are comparatively less astringent in mechanical terms than IBM approaches\cite{raos2011}. Indeed, DPD is a mass and momentum conserving algorithm that allows to model many-bodies embedded in a viscous fluid \cite{groot1997dissipative,espanol2017perspective,mukhi2021identifying}, even when geometries are complex \cite{xu2011dissipative,barai2016modeling}. {\it Quasi}-continuous DPD-approaches have been indeed revealed with a predictive capacity in describing certain rheological features of {\it Staphylococcus epidermidis} biofilms grown on a rheometer plate\cite{pavlovsky2013situ,IBM2}. By running DPD-simulations over long time and length scales\cite{raos2011}, the method has been also used to simulate biofilm formation on post-coated surfaces\cite{Balazs}, and the growth of two-dimensional biofilms under external flow \cite{Zhijie}.

More recently, some of us worked on a combined numerical and experimental study aimed at unravelling how hydrodynamic stress globally affects the mechanical features of {\it Pseudomonas fluorescens} biofilms \cite{jara2021}. The {\it P. fluorescens} biofilms were grown 
either under static or shaken conditions. Rheological measurements combined with confocal microscopy were further performed on the real biofilms. The experimental results showed the cultured {\it P. fluorescens} biofilms as capable of adapting to environmental conditions by tailoring their matrix microstructure to mechanical stress\cite{jara2021}. By considering prospective DPD simulations 
\textcolor{black}{allowing us to define the mesoscopic framework recapitulating principal ingredients from real biofilm behavior (polymer density, crosslinking degree, number of bacteria, etc.)}, those previous results suggested DPD-based rheological modelling with a forecasting potential. In particular, the DPD-simulations showed how much structural change was caused by an increased number of crosslinks in the EPS matrix \cite{jara2021}. In exploring the mechanical DPD-landscape as led by the living bacteria, however, the parameter space of the DPD method was not fully mapped, and more importantly, the viscoelastic behaviour of the biofilm to the applied shear frequencies was not studied in time domain.

\textcolor{black}{Beyond our previous work \cite{jara2021}, we here exploit novel coarse-grained DPD- modelling to numerically map rheological biofilm features in mechano-structural landscapes exploring nonlinear dynamic stress responses. The parameter space of biofilm viscoelasticity has been now mapped varying both mesh topology and structural compositions as represented by bonding interactions, such as the number of crosslinks between EPS polymers, the number of embedded bacteria, and the biofilm swelling amount of water molecules interacting along the DPD field. In order to check for dynamic (conservative / dissipative) features as structurally connected with the underlying biofilm mechanics, the rheological DPD- response has been now explored in time domain under variable strain spanning a broad range of shear amplitudes (both in linear and nonlinear regimes). The viscoelasticity predicted by the DPD-method has been further subject to experimental validation by measuring {\it in situ} responses of real {\it P. fluorescens} biofilms grown either under mild static conditions or intense shear stimuli. As a novel piece of retrospective physical forecasting on biofilm dynamics, our experimental findings with {\it P. fluorescens} biofilms have been discussed in the sight of the numerical outcomes obtained from the new DPD-simulations. }

The paper is organized as follows. We first describe the formal DPD-framework, its numerical set-up, and the validating experimental methods. The Results' section comprises an ample setting of numerical DPD simulations as mapping a broad rheological space at variable stress in two differentiated crosslinking topologies, either tightly (compact) or sparsely (homogeneous) crosslinked. The simulation results are then discussed on the sight of the experimental evidence accumulated on {\it P. fluorescens} biofilm rheology \cite{jara2021}, including new viscoelasticity measurements obtained at variable frequency under extreme conditions of growing, either statically unstressed or under shaking stresses. Finally, we summarize the conclusions.          

\section{Numerical and experimental methods}
 We present first the novel mesoscopic DPD-method used to simulate {\it in silico} the differently formed {\it P. fluorescens} biofilms as subject to variable shear stress spanning from {\it quasi}-static up to shaken conditions mimicking biofilm formation under hydrodynamic stress. The synthetic biofilms are prepared at periodic boundary configuration consisting of a colony of monodisperse bacteria immersed in a simulation box containing randomly distributed polymer and solvent. We also describe the experimental rationale designed for rheological measurements in biological biofilms at retrospective correspondence with the simulations on synthetic biofilms. 

\subsection{Mesoscopic DPD-simulations} {\color{black}
A given biofilm is modelled {\it in silico} as a mesoscopic system of interacting beads as recapitulating the three different components: bacteria, polymers and water.} First having built a polymer matrix as a mesh of rigid bonds with a fixed crosslinking we randomly insert monodisperse sized bacteria and finally hydrate the entire system with water molecules. Whilst 
  Raos \textit{et al}.\cite{GuidoRaos2006} dispersed the single bacteria as filler spherical particles, we build here rod-shaped bacteria by aggregating 440 simulation beads, each one with diameter $\sigma_b$, then brought together via harmonic spring interactions.
\begin{equation}
 U_{b}^{\rm bond}=\frac{K_b}{2}(r_{i,j} - \sigma_b)^2 \qquad,
\end{equation}
being $r_{i,j}$ the distance between bonded beads $i$ and $j$, and $K_b$ the harmonic coupling constant. The bacterium body is formed as a spherocylinder 
\textcolor{black}{whose central part is shaped as an empty cylinder of length $l_b/\sigma_b = 6$, while the extremes of the bacteria are shaped as spherical caps of radius $R_{ext}/\sigma_b = 4$. Both the central and extreme parts have an external radius
$R_{ext}/\sigma_b = 4$  and internal radius $R_{int}/\sigma_b = 2$ i.e. the bacterium membrane is
formed by three layers of particles}.

These rigid bacteria are immersed in a flexible mesh made of polymers (EPS) hydrated with water molecules. Water is represented by $N_{\rm solv}$ bead particles of type $s$. Whereas the EPS matrix consists of linear chains of $l$ beads bounded via an harmonic potential: 
\begin{equation}
 U_{p}^{\rm bond}=\frac{K_p}{2}(r_{i,j} - \sigma_p)^2 \qquad,
\end{equation}
where $r_{i,j}$ is the distance between two consecutive beads $i$ and $j$, $\sigma_p$ the equilibrium distance, and $K_p$ the coupling constant. The polymer beads of type $p$ can crosslink each other and with neighbouring bacteria (bacterium-polymer crosslinks). These crosslinking (CL) interactions have been also represented as harmonic potentials as established in Eqs. (1-2). 
Using the above structural approach all the bounding interactions are encoded inside the mesoscopic simulation box, namely, polymer-polymer (pp), polymer-bacterium (pb), polymer-solvent (ps), bacterium-bacterium (bb), bacterium-solvent (bs), solvent-solvent (ss).

Within the current simulation DPD-schema the total force between two particles $i$ and $j$ consists of three interactions, conservative, dissipative and random. The sum of the three DPD components captures long-range correlations induced by hydrodynamic interactions and also accounts for thermal fluctuations. The net force between two particles $i$ and $j$ (of type $\alpha$ and $\beta$) can be expressed as the sum of the conservative force $\vec{F}^C_{i,j}$, plus the dissipative force $ \vec{F}^D_{i,j}$ and the random force $\vec{F}^R_{i,j}$. The conservative force corresponds to $\vec{F}^C_{i,j} = B_{\alpha,\beta}w(r)\hat{r}_{i,j}$, for $ r<r_c $, where $r_c$ is a cut-off distance beyond which all these terms vanish;  $\hat{r}_{i,j}$ is the unit vector along the direction of  $\vec{r}_i -\vec{r}_j$, indicating the positions of  particles $i$ and $j$, respectively; $B_{\alpha,\beta}$ is the amplitude of the conservative force between particles of type $\alpha$ and $\beta$. The dissipative force is represented as $\vec{F}^D_{i,j} = -\gamma w^2(r)\left(\hat{r}_{i,j}\cdotp \hat{v}_{i,j}\right)\hat{r}_{i,j}$, where $\vec{v}_{i,j}\equiv\vec{v}_i-\vec{v}_j$ is the vector difference between the  velocity $\vec{v}_i$ of particles $i$, and the velocity $\vec{v}_j$ of  particle $j$. Here, $\gamma$ is the friction coefficient. The random force can be computed as $\vec{F}^R_{i,j} = w(r) \left(\dfrac{2k_BT\gamma}{dt}\right)^{1/2} \Theta \, \, \, \hat{r}_{i,j} ,  w(r) = 1-r/r_c$, being $w(r)$ a weighting factor varying from 0 to 1.  $\Theta$ is a Gaussian random noise with zero mean and unit variance; $dt$  the integration time, $k_B$ the Boltzmann constant, and $T$ the absolute temperature. {\color{black} This random force essentially captures the microscopic Newtonian friction as balanced against the fluctuating impulsion by the thermal energy transferred from the environment}.

By using internal units we set $r_c=1$ for all DPD interactions, $\sigma_b=\sigma_p=0.5$, $\gamma=4.5$, $K_b=K_p=30$. The time step is set to $dt=0.05$, although we test our results also against $dt=0.005$.
To convert dimensionless units in physical units, we consider $k_BT=4.11\times 10^{-21}$ J as the characteristic energy scale, and  we set the length scale to the longest dimension of one \textit{P. fluorescens} bacterium,  approximately $\sim 1.5 \mu$m.  
\textcolor{black}{Since the size of one bacterium is about 1.5 $\mu m$, and one bacterium os about 10 beads long, the diameter of one bead  ($\sigma_b$) corresponds to about 100 nm. Having assumed $\sigma_b=\sigma_p$, the size of a single polymer in the experiments  (about 100 nm) would correspond to a single bead. For this reason, in our simulations we assume that  polymer chains inserted in the system represent a network of connected polymers, comparable in size with the bacteria.}
Moreover, since the mass of a single bacterium is $\sim 10^{-15}$ Kg, the mass of a single particle is set to $m_u \sim 2.3\times 10^{-18}$ Kg. The  time scale is derived accordingly as $\tau_{\rm intrinsic} = \sqrt{m_ul_u^2/k_BT}\sim 5\times10^{-6}$s. 

\subsection{Synthetic biofilm: simulation scheme}

We prepare configurations of bulky $32\times32\times32$ biofilm boxes which are evolved in time via a constant-volume DPD dynamics performed using an open source LAMMPS-integrator package imposing periodic boundary conditions\cite{lammps}. To study rheological features the parameter space is explored as follows: a number of bacteria $N_b$ ranging from 100 to 200; a number of polymers $N_p$ from 80 to 200 
\color{black}{(of polymer lengths $L_p=100$)}; a number of solvent particles $N_s$ from 20,000 to 50,000. 
All parameters $N_p$, $N_s$ and $N_b$ have been chosen to fulfill the DPD condition for the density\cite{dpd,groot1997dissipative}; this is $(N_p + N_b + N_s ) / ( 32 \times 32 \times 32) > 3$. To guarantee bacteria and polymers being properly hydrated, we choose the amplitudes of the solvent interactions smaller than any others. In particular, we choose $B_{s,s}=B_{s,b}=B_{s,p}=25$ and $B_{b,b}=B_{p,p}=B_{p,b}=30$. 
\textcolor{black}{When varying the water content, to study the effect on the rheology when  changing the biofilm solvation, we accordingly changed $N_b$ and $N_p$  to always keep the overall system’s density larger than 3}.

\textcolor{black}{In order to understand whether our polymer suspension is in a dilute or semi-dilute regime, we compute the average polymer concentration of our simulations and relate it to the overlap concentration. We estimate the polymer overlap concentration as  $c^* \sim N_p/L_p^3$, where $N_p$ is the number of monomers in one polymer and $L_p$ the  polymer length. In a theta-solvent, we can approximate  $L_p \sim N_p$, leading to $c^* \sim 1/N_p^2 = 0.0001$ (being $N_p \sim 100$).  Given that we insert about 100 polymers, each one of length 100, the average polymer concentration is about 0.3, that is above the overlap concentration. Thus, we are dealing with a semi-dilute polymer system.}

To prepare initial configurations, we randomly locate $N_b$ bacteria and $N_p$ polymers in the presence of$N_s$ water molecules and then equilibrate without polymer cross-links (CL) for $\sim 8\times 10^6$ time steps. Next, we turn on the crosslinks within the polymer network. For any choice of $N_b$ and $N_p$, we form a fixed number of crosslinks ranging from 2,300 up to 5,400. 
To create the network of crosslinked polymers and bacteria, we randomly place $N_{\rm CL}$ newly formed harmonic bonds: 1) between $b$ and $p$ particles i.e., bacterium-polymer crosslinks, and 2) between two $p$ particles belonging or not to the same polymer chain i.e., polymer-polymer crosslinks. 
To avoid the formation of artificial aggregates of $p$ particles, we forbid crosslinks between nearest-neighbour beads belonging to the same polymer chain. Thus, we choose on the one hand to inhibit crosslinks between the ten closest neighbouring particles belonging to the same chain. The resulting homogeneous topology is referred to as $T_A$ (with a probability for each chain monomer to crosslink $p_A = 1/10$). On the other hand, we choose to allow crosslinks between particles that are more than three neighbors apart. We refer to the resulting polymer compact topology as $T_B$ (with a three-fold higher crosslinking probability than $T_A$ i.e., $p_B = 1/3 > p_A$). Unless specified otherwise all the presented results will refer to the topology $T_A$, considered to be sufficiently open to allow for bacterial reorganizations. To prevent excess formation of crosslinks per particle, which would result in a globule-like cluster of particles, we assume that any particle of type $b$ can form at most one crosslink, while any $p$ particle can form at most two crosslinks. After crosslink formation a second equilibration run of $\sim 10^6$ time steps is performed to relax the polymer-bacteria network. 

\begin{figure}[h!]
\centering
  \includegraphics[width=0.25\textwidth]{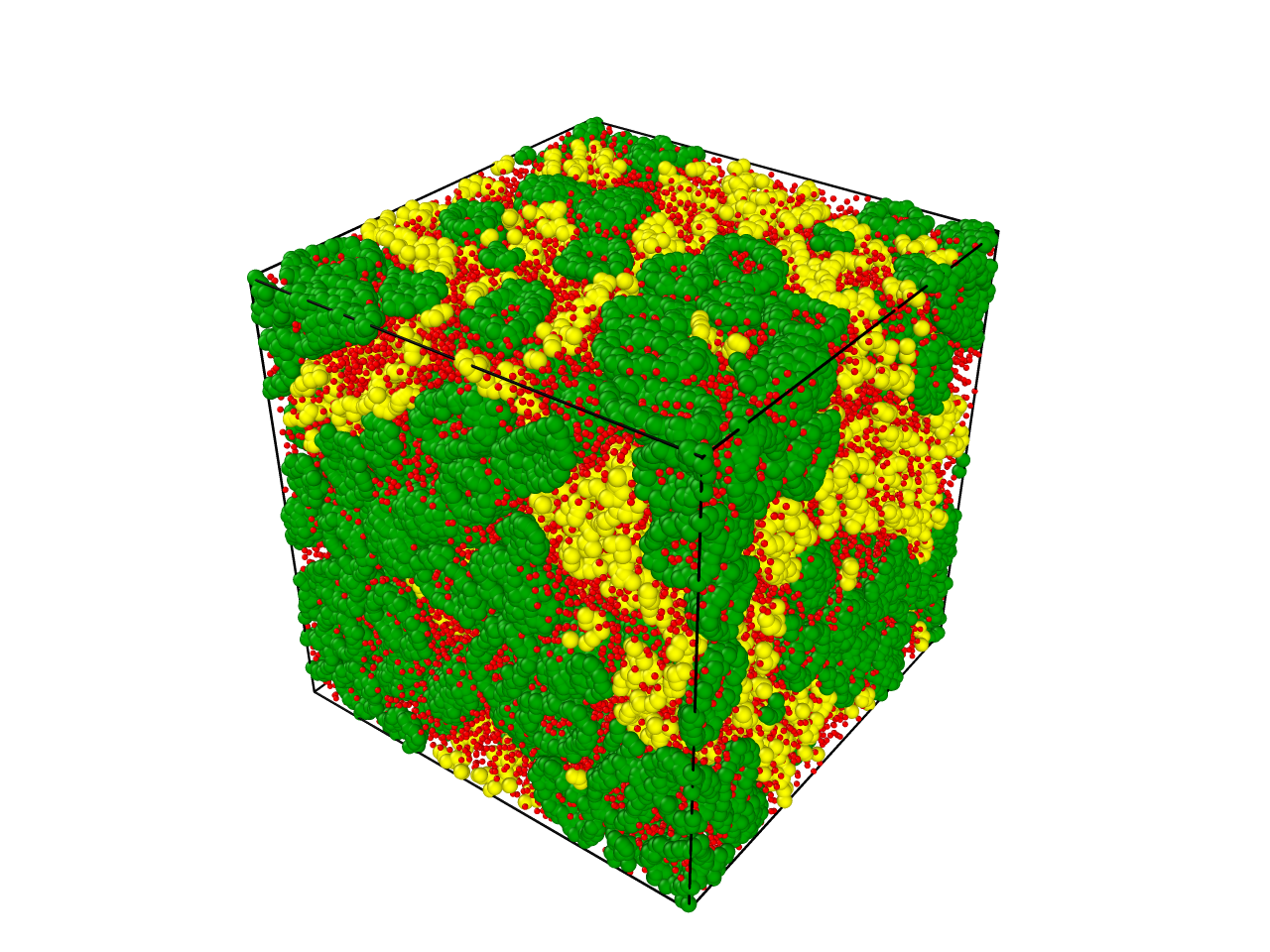}
 \includegraphics[width=0.22\textwidth]{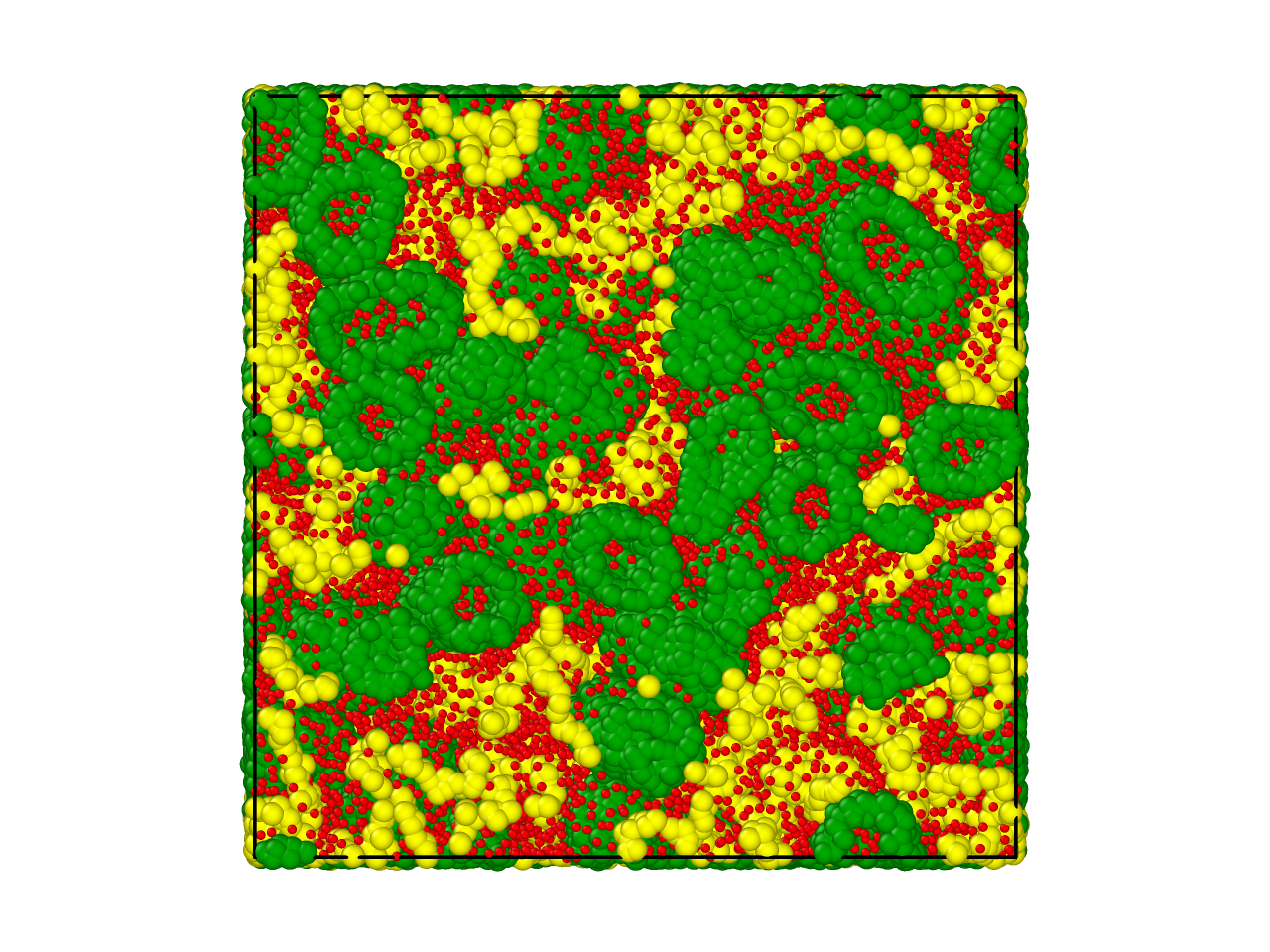}
 \caption{{\color{black} Synthetic DPD-biofilms in the homogeneous $T_A$ topology: Left panel: Simulated structure after numerical equilibration. Snapshot of the simulation box containing 184 bacteria (in green), 80 polymers (in yellow) and 35,000 solvent particles (in red). 
 Right: Cross-section of the same system.}}
 \label{fig:model}
\end{figure}

Figure \ref{fig:model} shows an example of equilibrated configuration corresponding to the homogeneous $T_A$ topology considered undeformed (in the absence of shear stress).

\noindent
\subsection{Canonical $T_A$ topology: pair distribution function}

\begin{figure}[h!]
    \centering
    \includegraphics[width=0.47\textwidth]{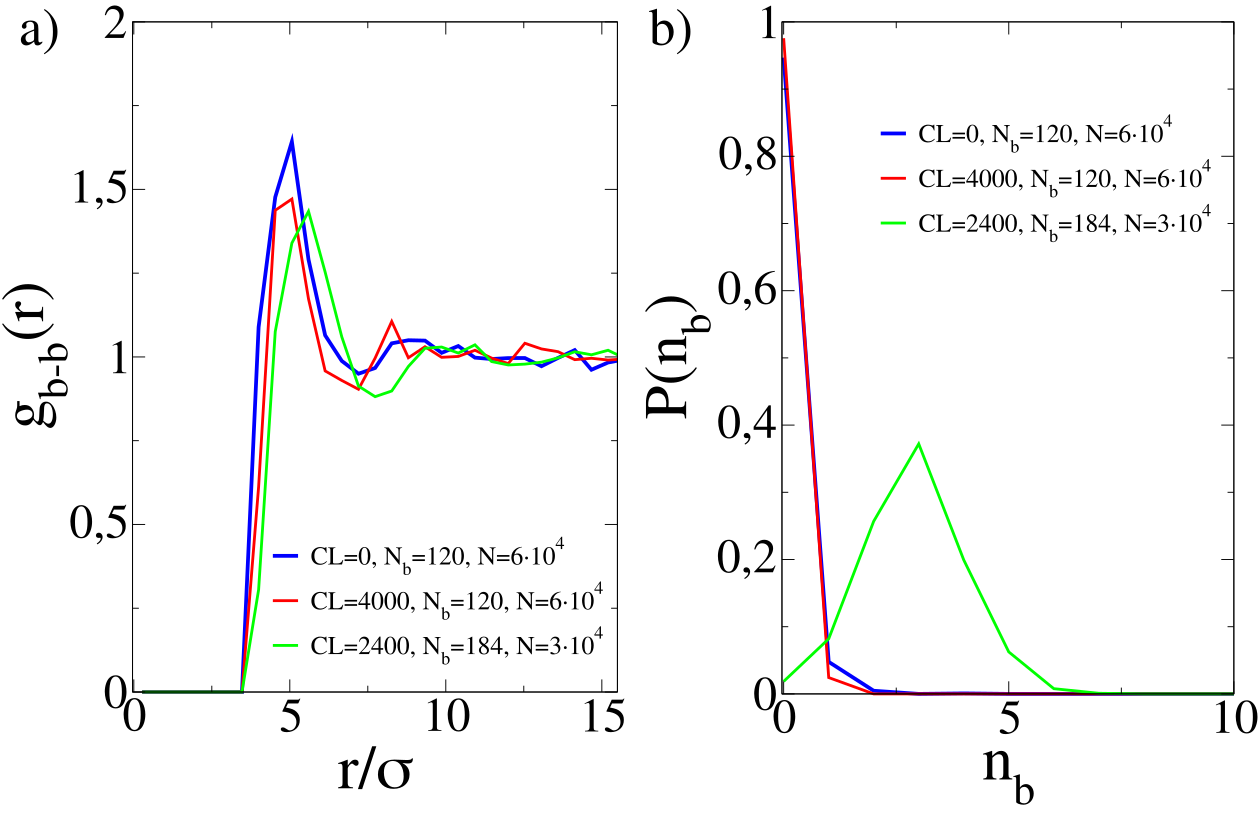}
    \caption{(a) Radial pair distribution function for the center of mass of bacteria for three sets of parameters: CL=0, $N_b=120$, $N_p=80$, $L_p=100$, $N_s=60000$ (blue); CL=2400, $N_b=184$, $N_p=80$, $L_p=100$, $N_s=30000$ (green) and CL=4000, $N_b=120$, $N_p=80$, $L_p=100$, $N_s=60000$ (red). (b) Probability distribution function for number of bacteria in box slices of size $L/4$ (system parameters as in panel (a)).
    }
    \label{fig:gdrpdf}
\end{figure}

As expected, the homogeneous $T_A-$biofilm is characterised by an homogeneous distribution of bacteria within a weakly crosslinked matrix as characterized by the radial distribution function $g_{b-b}(r)$ computed on the bacteria's center of mass (Fig. \ref{fig:gdrpdf}a). 

All the radial distribution plots exhibit a marked peak corresponding to nearest-neighbor correlations at short distances $r/\sigma\sim 5$, while approaching $g_{b-b}(r)\sim1$ at longer distances, as expected in homogeneous liquids. Regardless of the number of bacteria $N_b$, the number of polymers $N_p$, the number of crosslinks CL and the level of hydration $N_s$, all curves representing the $g_{b-b}(r)$ show exactly the same behaviour (differently colored curves in Fig. \ref{fig:gdrpdf}a). As representative configurations, the blue and red curves reported in Fig. \ref{fig:gdrpdf}a refer to systems with different CL numbers, whereas the  green and blue curves refer to systems who differ in $N_b$, CL and $N_s$.

In order to test the occurrence of bacteria aggregation in the initial sample, we look for the appearance  of phase separation (which would result in clustering of bacteria). For this purpose, we  compute  the density distribution of  bacteria's center of mass, shown in Fig. \ref{fig:gdrpdf}b. To perform this calculation, we divide the box volume in cubic cells of size $L/4$ and compute the number of bacteria per cell. Finally, the distribution is computed over multiple cells taken from several independent configurations. The  blue and red curves refer to systems with a lower number of bacteria, resulting in averaging mostly empty cells, while the green curve refers to a system with a larger number of bacteria, resulting in a non-vanishing average density. Independently on the the number of bacteria $N_b$, of polymers $N_p$, of crosslinks CL and of the level of hydration $N_s$, we  observe  a single-peak distribution. This underlines the fact that the system is locally structured but not phase separated (in agreement with the expected biofilm homogeneity in the weakly crosslinked $T_A$ topology). 
\textcolor{black}{Even though the snapshots in Figure 1 seem to suggest that biofilms are spatially inhomogeneous, neither of the two procedures (TA or TB, data not shown) lead to a phase separated system. Work is in progress to build a model characterised by a clear spatial inhomogeneity, based on a combination of polymers of different length, together with a biased bacterial aggregation.}

\textcolor{black}{For a fixed number of bacteria, polymers and water content, we have estimated the average distance between two connected beads in the polymer mesh and demonstrated that it is a decreasing function of the total number of crosslinks, as expected. Since the larger the total number of crosslinks, the larger the number of crosslinks between polymers, the smaller the average distance between two connected beads in the polymer mesh. When the number of crosslinks is larger than 2000, the average distance between two connected beads in the polymer mesh does not vary considerably (data not shown). Throgout the rest of the work, we will only consider biofilm matrix in the $T_A$ topology and with a number of crosslinks larger than 2000.}

\subsection{Numerical DPD-rheology}
To study the rheological properties of a given model biofilm, we compute the shear modulus by monitoring the shear stress response $\sigma_{xy}(t)$ as resulting from an imposed sinusoidal deformation of the simulation box\cite{raos2011,jara2021}. We apply an external oscillatory shear deformation with time frequency $\nu$ along the $X$--$Y$ plane by changing the box size $L$ according to \begin{equation}
L (t)= L_0 + A\sin(2\pi \nu t) \qquad,
\label{eq:shear_zsize}
\end{equation}
being $L_0=32$ the initial box size and $A$ the oscillation's amplitude.
According to  Raos \textit{et al}.  \cite{GuidoRaos2006}, the resulting stress can be calculated by fitting the $xy$ component of the shear stress with: 
\begin{equation}
 \sigma_{xy}(t)=\sigma'\cdot \sin(2\pi \nu t) + \sigma''\cdot \cos(2\pi \nu t) 
 \label{eq:sigma_xy}
\end{equation} 
with $\sigma'$ and $\sigma''$  corresponding to the the in-phase and out-of-phase components of the complex shear modulus $G$, whose components are \begin{equation}
G' = \frac{\sigma'}{(A/L_0)} \qquad G'' = \frac{\sigma''}{(A/L_0)}
\end{equation}
being $G'$ the storage modulus and $G''$ the loss modulus, respectively.
Whereas rheological properties of solids are characterized by a finite storage modulus $(G^{'}>0$), fluid viscous materials are characterized by near zero storage $(G^{'} \sim 0$) and large losses $(G^{''}>0$)\cite{donlan2001biofilms,alonci2018injectable,findley2013creep}.

When shearing the DPD-system, we choose to vary the amplitude $A$ between 4 and 36 corresponding to a maximal deformation strain ($u=A/L_0$) of up to the 112\% of the box size. We mainly focus on data obtained for $A=24$ ($u=$ 75\%), since the stress-strain curve behaves almost linear nearby but departs from linearity for larger deformations. We herein find the response stronger and the noise lower when varying system parameters ($N_b$, $N_p$, ...). Under shearing conditions, particle velocities are remapped every time they cross periodic boundaries.

\color{black}{Having performed preliminary numerical experiments with a simulation box of $16 \times 16 \times 16$, we obtained the same rheological behaviour as with a larger system of $32\times32\times32$. Therefore, we chose   the larger box  to study the bulk rheological  features of a bacterial biofilm.}
 We choose a period $T\equiv\frac{2\pi}{\nu}$ ranging between  $2 \times 10$ up to $2 \times 10^5$. 
At a time step of $dt=0.01$, a period of $T=200$ corresponds to 2000 time steps that is within the total lifespan of the simulation runs. For the considered frequency, ranging from a value of $3\times10^{-5}$ to a value of $0.3$ (in internal units), we run the simulations completing at least 3 box-shearing oscillations. Once the numerical simulations of shearing have been performed in internal units, the rheological results can be converted into physical units by multiplying the shear stresses inside the simulation box into physical pressures in each $XY$-plane wherein the shear strain is a constant. For the considered simulations the conversion factor is $N_{XY} m_u/(l_u \tau_u) = 4.1$ kPa per internal unit of pressure as considered in each shear plane containing $N_{XY} = 32 \times 32 = 1,024$ particles.

\noindent
\subsection{Viscoelasticity regimes} 
As a rule of thumb on the proposed DPD-rheology, we performed simulations in synthetic biofilms with an intermediate number of particles interacting {\color{black} in rigid realizations under dominant chain crosslinking} in the canonical $T_A$ topology. Figure 3 shows typical stress-strain plots obtained for considered medium and high crosslinking (respectively CL = 3600, 5300), and extremal values of the shear frequency. 

\begin{figure}[h!]
    \centering
    \includegraphics[width=0.45\textwidth]{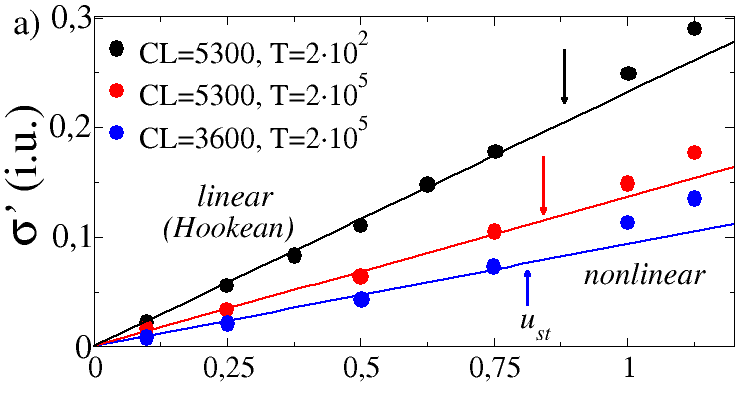}
    \includegraphics[width=0.45\textwidth]{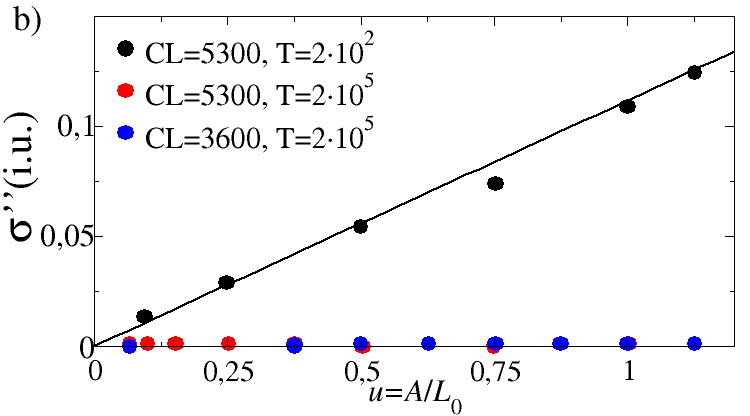}
    \caption{Rheological DPD-response of synthetic biofilms. In plane stress $\sigma^{‘}$ (panel a) and $\sigma^{‘‘}$ (panel b) as a function of $A/L_0$ for three different biofilms characterised by 5300 crosslinks  and $T=200$ (black symbols),  5300 crosslinks  and $T=2 \times 10^5$ (red symbols), 3600 crosslinks  and $T=2 \times 10^5$ (blue symbols). The straight lines are needed to underline the regime where we consider the shear to be linear. All  data refer to $N_b=150$, $N_p=80$ and $N_s=45000$
    }
\end{figure}

From a phenomenological point of view, the computed stresses are characterized by two regimes at dependence of the applied strain on the synthetic DPD-biofilms ($u=A/L_0$), {\color{black} which imposes the stress response at dependence of the prescribed degree of crosslinking in the simulated topologies.} 

On the one hand, for the in-phase stresses we detect (Fig. 3a): a) A Hookean elastic regime at low strain below a strengthening point ($u < u_{st}$); here, stress and strain are linearly related until the onset for dynamic rigidization (nearby $u_{st}$). b) A nonlinear regime beyond the strengthening point ($u > u_{st}$); here, the system strengthens under nonlinear stress thus becoming effectively stiffer than in the Hookean regime. Very relevantly, the higher crosslinks' number the strengthener the resulting synthetic biofilms. 
 As expected for rigid networks made of semi-flexible polymers with permanent crosslinks \cite{aufderhorst2019stiffening}, the most rigid synthetic realization arises from the densest crosslinked mesh strained at the highest frequency (at short deformation period $T= 200$).

 On the other hand, the out-of-phase stresses show the viscous losses as a dissipative frictional response (Fig. 3b). The presence of finite (non-zero) viscous shear stresses is only detected for the most rigid realizations with the highest density of crosslinks strained at high frequency (black symbols); no viscous stress is detected at low frequency (red symbols), nor at low density of crosslinks (blue symbols). Therefore, we identify the densely crosslinked as a viscoelastic material. Whereas a high systemic shear rigidity makes the elastic response to emerge solid-like ($\sigma$' $\sim G' u$ with $G'>> 0$), the finite dynamic viscosity imposed by the DPD algorithm makes the frictional losses to effectively emerge only at high frequency (i.e., $\sigma$'' $\sim G'' u $ being $G''=\omega\eta$ the loss modulus as determined by the shear viscosity $\eta$). Otherwise, the synthetic system behaves inviscid ($\sigma$'' $\sim 0 $).

\noindent
\subsection{{In vitro biofilm formation}} 
The strain \textit{P. fluorescens} B52, originally isolated from raw milk \cite{Richardson1978}, is used as a model microorganism. Overnight precultures and cultures are incubated at 20$^\circ$C under continuous orbital shaking (80 rpm) in tubes containing 10 ml Trypticase Soy Broth (TSB, Oxoid). Cells are recovered by centrifugation at 4,000\textit{g} for 10 min (Rotor SA-600; Sorvall RC-5B-Refrigerated Superspeed Centrifuge, DuPont Instruments)
and washed twice with sterile medium. Cellular suspensions at OD$_{600}$ are first adjusted to 0.12 (equivalent to 10$^8$ cfu/ml) and then diluted  to start the experiments at  10$^4$ cfu/ml. Biofilms are grown on borosilicate glass surfaces ($20\times 20$ cm) as adhesion substrates. Five glass plates are held vertically into the sections of a tempered glass separating chamber, provided with a lid. The whole system is heat-sterilized as a unit before aseptically introducing 2 ml of the inoculated culture medium. To check the effect of hydrodynamic stress on biofilm mechanical properties, incubation is carried out for 96 h at 20$^\circ$C both in an orbital shaker at 80 rpm (shaken sample) and statically (non-shaken sample). For biofilm recovery, plates are aseptically withdrawn, rinsed with sterile saline to eliminate weakly attached cells, and then scraped to remove the attached biomass (cells + matrix) from both sides of the plates. For rheological measurements, the biofilm material is casted every 24 h to be directly poured onto the rheometer plate. Experiments are run in triplicate.

\noindent
\subsection{Experimental rheology} 
The biofilm's viscoelastic response is experimentally determined in a hybrid rheometer under oscillatory shear stress-control (Discovery HR-2, TA Instruments), using a cone-plate geometry (40 mm diameter) and a Peltier element to control temperature\cite{jara2021}. Triplicate measurements are performed at a 1 mm gap between the Peltier surface and the cone-plate tool (TA instruments), where a sinusoidal shear strain $u(t)$ of amplitude $u_0$ is performed at a frequency $\omega = 2\pi\nu$, i.e., $u(t) = u_0$ $sin (\omega t)$. The shear deformations are considered at variable angular frequency ($\omega=2\pi\nu$). The lower frequencies are restricted by the extremely long readout times compromising sample stability ($\nu_{min} = 10^{-2} Hz$). The practicable frequency window is upper limited by inertia ($\nu_{max} = 100 Hz$); higher oscillation frequencies are not usually considered to be affected by artifacts in a blind region dominated by inertia. Measurements are performed at low strain amplitudes (typically $A= 1\%$), for which the stress responses are found practically linear. The shear stress exerted by the biofilm in the linear regime is monitored as $\sigma(t) = G^{*} u(t)$, where $G^{*}$ is the viscoelastic modulus $G^{*} = G^{'} + i G^{''}$. The storage modulus accounts for Hookean shear rigidity ($G' \sim G_0$), and the loss modulus for Newtonian viscous friction ($G''=\eta\omega$; at constant shear viscosity $\eta$). 

\section{Results}
Our numerical simulations focus on mechanical measurements covering linear and nonlinear regimes of predictive viscoelasticity in synthetic DPD-biofilms. As guided by the preliminary simulations performed to determine the viscoelasticity regimes (see Fig. 3), we build upon our numerical setup for near-conservative DPD-rheology by keeping fixed $N_b = 150$, $N_p = 80$, and $N_s = 45000$ (corresponding to a well populated biofilm colony).
 Hereinafter,  all data refer to a low frequency deformation that minimises the frictional losses  (at period $T = 2$ x $10^5$). 
  Simulation runs were performed as a function of the shear amplitude $u=A/L_0$, for different number of polymer-polymer, or polymer-bacteria crosslinks with topology either $T_A$ or $T_B$. 
  {\color{black}The DPD-realizations here performed actually correspond to relatively lower degrees of crosslinking as above considered in Fig. 3. They should capture the fuzzy structure expected for realistic EPS-networks in which the embedded particles are able to explore an ample configurational space even under relatively small amplitudes of the shear field externally applied.}

\subsection{Synthetic mechanical response under increasing shear deformation: influence of biofilm topology}

\begin{figure}[h!]
\includegraphics[width=0.5\textwidth]{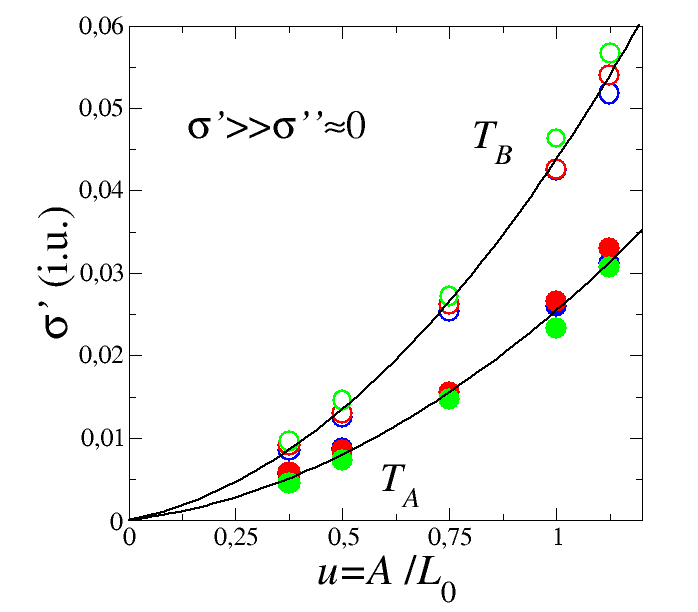}
 \caption{Conservative shear response of synthetic DPD-biofilms under shear strain. The in-phase stress $\sigma^{‘}$ is represented as a function of $A/L_0$ for three different biofilms characterised each by a different topology: $T_A$ (solid circles) and $T_B$ (hollow circles); in internal units (i.u.). The color code indicates different numbers of polymer-polymer (pp), polymer-bacteria (pb) cross-links: $CL_{pp}=200$ / $CL_{pb}=1,400$ (blue); $CL_{pp} = 400$ / $CL_{pb}=1,200$ (red) and $CL_{pp}=600$ / $CL_{pb}=1,000$ (green). All  data refer to $N_b=150$, $N_p=80$, $N_s=45,000$ and $T=2\times10^5$. The straight lines represent parabolic fits to simulation data (see main text).} 
 \label{figure3} 
 \end{figure}

In order to characterize the parameter space of dynamical responses to shear as corresponding to the considered crosslinking topologies ($T_A$ and $T_B$), we compute the mechanical stress as a function of the shear strain for biofilms as those prepared in the numerical methods' section (keeping fixed $N_b=150$, $N_p=80$, $T=2 \times 10^5$, $l=100$ and the total number of crosslinks $CL_{pp} + CL_{pb}=1,600$).  The results from these DPD-simulations are reported in Fig. \ref{figure3} (symbols corresponding to systemically different DPD-realizations). The elastic (in-phase) stress shows a Hookean limit of linear response at low shear deformations ($u=u_C\sim 50 \%$), followed by a nonlinear trend towards higher stresses at larger deformations ($u >> 75 \%$). Differences appear when looking at the magnitude of $G^{'}$ if considering biofilms containing crosslinks between particles within the same polymer chain. If they are more than three neighbors apart as in the compact $T_B$ topology the biofilm is always more rigid (Fig. 4; hollow symbols), than when containing crosslinks between particles that are at least ten neighbors apart i.e., the homogeneous topology $T_A$ (solid symbols). The performed DPD-simulations show that crosslinks between particles that are closer along the same polymer chain lead to stiffer biofilms: this might be the consequence of aggregation induced in the polymer chains. Under shear strain the data become clearly grouped as corresponding to the studied topologies; these are: $T_A$) The structurally homogeneous $T_A$-topology, showing low stress and broad linear regime under deformation; $T_B$) The structurally compact $T_B$-topology, much stiffer, responding hence with higher stresses and ampler non-linearity. We superpose parabolic fits as straight lines to indicate the amplitude range where we can consider the quadratic shear response as limited by the leading Hookean component. The viscous (out-of-phase) stresses remain practically vanishing in these settings which behave practically frictionless ($\sigma^{''} \sim 0$; data not shown). {\color{black} In both topologies $T_A$ and $T_B$ sheared under large amplitude deformation ($u \geq u_{st}$), the current DPD-simulations evidence nonlinear strengthening without increasing frictional losses as due to conservative elongational chain ordering templated between the permanent crosslinks of a rigid meshwork.}  

\textcolor{black}{Differently to our previous experimental results with {\it P. fluorescencs} biofilms \cite{jara2021}, and unlike rheological biofilm data reporting on nonlinear softening e.g., in {\it S. epidermis, S. mutans, etc.} (see Ref. \cite{jana2020nonlinear} for a review), the permanent crosslinking considered by our DPD-simulations predicts biofilm hardening as expected for polymer chains becoming rigidly ordered under stress. Such class of strengthening stresses have been observed in modified biofilms composed by variants of {\it P. aeruginosa} able to overproduce rigid EPS polysaccharides \cite{waters2014}. This confirms our "stiff" DPD-simulation framework too rigid to capture collective relaxations from mobile crosslinks and sliding entanglements as existing in the real biofilms. Despite these obvious limitations, we resume our analysis on the possibilities, capacities and strengths of the current DPD-simulations schema as based on permanent crosslinks.}

\subsection{Effective (linear and nonlinear) shear rigidity} 
We further calculate the effective rigidness of the studied DPD- biofilms defined as the apparent modulus for elastic storage i.e., $G'=\sigma {'}/u$ (see Eq. 5; left). {\color{black} This effective parameter measures the apparent rigidity as recapitulated in a shear modulus for elastic storage. To calculate the effective rigidity as a function of strain $G'(u)$, we use raw simulation data on the stress-strain response (Fig. 4).} Figure 5 plots simulation data for the apparent storage modulus $G'$ increasing non-linearly with the shear amplitude (found larger in magnitude for the stiffest $T_B$ biofilm than for the canonical $T_A$ topology). The linear rigidity modulus is given by the Hookean limiting intercept at zero-strain ($G_0$ at $u =0$).\\

 \begin{figure}[h!]
 \includegraphics[width=0.5 \textwidth]{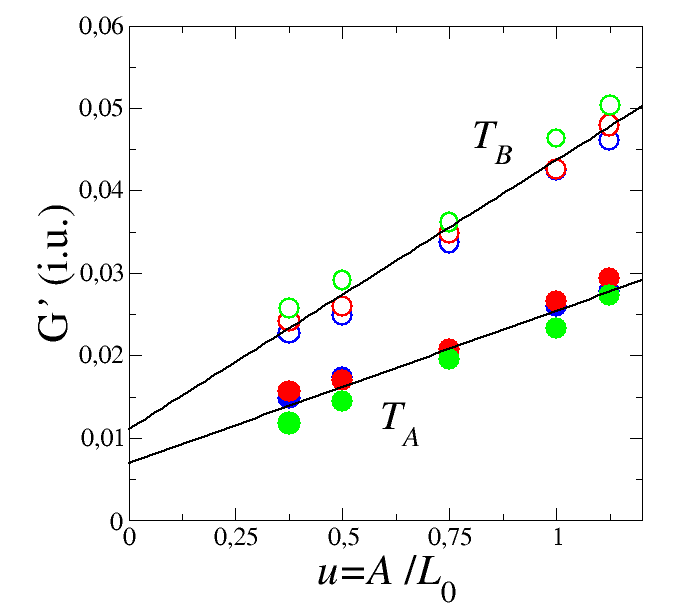} 
 \caption{Storage modulus $G'$ as a function of strain for the topology $T_A$ (filled squares) and $T_B$ (empty circles); in internal units (i.u.). The color code indicates different numbers of polymer-polymer (pp), polymer-bacteria (pb) cross-links: $CL_{pp}=200$ / $CL_{pb}=1400$ (blue); $CL_{pp}=400$ / $CL_{pb}=1200$ (red) and $CL_{pp} =600$/ $CL_{pb}=1000$ (green). All data refer to $N_b=150$, $N_p=80$, $N_s=45000$ and $T=2\times10^5$. The straight lines represent linear fits to simulation data (see main text).}
\label{stress_tropology}  
\end{figure}
 
The effective storage modulus increases with the strain-amplitude in both topologies (being larger in magnitude for the more compact $T_B$). This is the sort of behavior expected for semiflexible newtworks undergoing stress strengthening as here simulated \cite{aufderhorst2019stiffening}. We observe in both cases that the shear rigidity is not too much affected by the ratio between polymer-polymer ($pp$) versus polymer-bacteria ($pb$) crosslinks, as long as the total number of CL remains constant in the simulations. To calculate the Hookean moduli ($G_0$), and the nonlinear amplitudes ($g_1$), we exploit the phenomenological dependency $G'=\sigma {'}/u \sim G_0 + g_1 u + ...$ (see linear fits in Fig. 5). The fitting parameters are collected in Table I.  

\begin{table}[h!]
\begin{tabular}{c|c|c|}
\cline{2-3}
                            & $G_0$ (i.u.)      & $g_1$ (i.u.)      \\ \hline
\multicolumn{1}{|l|}{$T_A$} & $0.007 \pm 0.001$ & $0.019 \pm 0.001$ \\ \hline
\multicolumn{1}{|l|}{$T_B$} & $0.011 \pm 0.001$ & $0.033 \pm 0.002$ \\ \hline
\end{tabular}
\caption{Characteristic parameters of shear rigidity calculated from the storage modulus in the studied DPD- topologies. $G_0$ is the Hookean shear rigidity, and $g_1$ the nonlinear amplitude that describes biofilm strengthening under stress. The internal units of rigidness (i.u.) can be converted into physical units of pressure by the factor 1 int. stress unit $ = N_{XY} m_u / (l_u \tau_u) = 4.1$ kPa for $N_{XY}=32 \times 32$ particles in each shear plane.}
\end{table}

In both topologies ($T_A$ and $T_B$), we compute the Hookean modulii with values around $G_0 \sim 0.01\sim 1 $ kPa (corresponding to a simulation box with $N=32,768$ particles inside); they estimate the mechanical rigidness of the synthetic biofilms in qualitative agreement with previous experiments performed with the \textit{P. fluorescens} biofilms studied in the linear rheological regime\cite{jara2021}. The fitted values found for the nonlinear amplitudes $g_1$ 's reveal strong biofilm strengthening under stress which is however not observed in the experimental window explored in the previous paper \cite{jara2021}. The current simulations show the effective DPD-rigidities increasing by more than a 100\% for $T_A$, and 300\% for $T_B$ (considered both at 100\% strain i.e., at $u=1$; see Fig. 5). Interestingly, the rigidness ratios indicate the relative softness of the homogeneous topology $T_A$ with respect to the compact one $T_B$ i.e., $G_0^{(A)}/G_0^{(B)} \sim g_1^{(A)}/g_1^{(B)} \sim 0.6$. These mechanical ratios are hopefully determined by structural crosslinking probabilities relative to both simulated topologies i.e., $1-p_A/p_B \sim 0.6$ as encoded in the DPD- method (see Numerical Methods). Therefore, and in order to accommodate enough structural flexibility in the  DPD- biofilms numerically simulated in the following, we will only deal with the homogeneous meshworks modelled at inhibiting crosslinks between the ten closest neighbouring particles belonging to the same chain (canonical $T_A$ topology).
 
 \begin{figure*}[t]
 \makebox[\textwidth][c]{\includegraphics[width=0.87\textwidth]{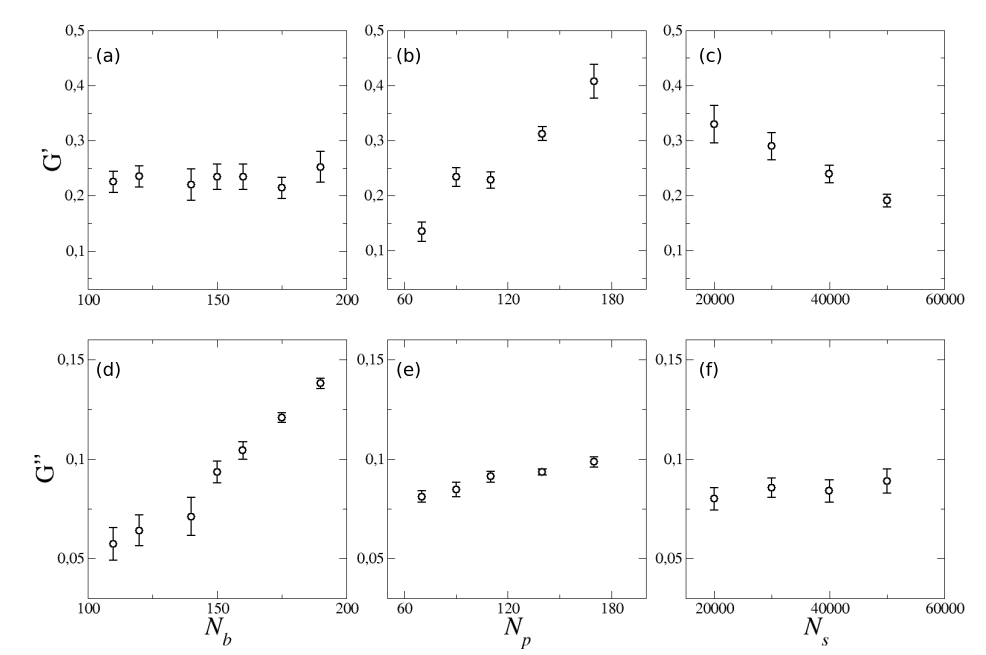}}%
 \caption{Stress components $G'$ and $G''$ for $A=75\%$, $CL=2300$ and $T=200$ as function of (a,d) the number of bacteria $N_b$; (b,e) the number of polymers $N_p$ and solvent (c,f).}
 \label{fig:Loss_Storage_Modulus}
 \end{figure*}
 
 \subsection{Tuning structural parameters into synthetic DPD-modeled biofilm viscoelasticity}
 \noindent
To analyze how far structural parameters such as $N_b$, $N_p$, and $N_s$ affect biofilm viscoelasticity, we consider the homogeneous DPD-biofilms created with the canonical $T_A$ crosslinking topology (i.e. by inhibiting crosslinks between the ten closest neighbours in a polymer chain). We study stress responses from those flexible $T_A$ biofilms designed to be a flexible meshwork within a dissipative dynamics (see Numerical Methods for details on DPD simulations). Figure \ref{fig:Loss_Storage_Modulus} shows the dependence of the mechanical properties on $N_b$, $N_p$ and $N_s$ as implemented in the canonical $T_A$ topology. We set an intermediate number of crosslinks ($CL=2300$), and intermediate values of $N_p=80$, $l_p=100$, $N_s=35000$ within the explored range. Having prepared a DPD-biofilm with an equilibrated $T_A$ topology, we apply a deformation amplitude of $A=75$ $\%$ and a deformation period of $T=200$, both comprised well within a viscoelastic response entering the onset of nonlinearity. {\color{black} Significantly, we aim at mapping the effective response of the simulated DPD-biofilms in broad ranges of rheological behaviour making the compositional effects to emerge in the apparent viscolastic moduli $G'$ and $G''$ (as calculted from Eqs. (5)).} These choices on the $T_A$ topology are dictated by the fact that they correspond to a significant response of $G''$ as we have observed at high frequency, differently from the vanishing losses as happening at lower frequencies. Three different structural biofilm variations will be relevantly considered as referred to changes in the number of bacteria ($N_b$), of polymer matrix crosslinks ($N_p$), and of water (solvent) molecules constituting the aqueous environment ($N_s$). These constitutional effects are discussed below.

{\it Bacterial content}. To check whether the shear stress depends on the number of bacterial cells ($N_b$), we have computed $G'$ (in-phase response to strain), and $G''$ (out-of-phase response) as a function of $N_b$ while keeping all other parameters fixed (Fig. \ref{fig:Loss_Storage_Modulus}-panel a and d). On the one hand, simulations show that the in-phase response (accounting for the storage modulus $G'$) remains constant throughout the $N_b$ range (panel a). This result is in agreement with previously reported data \cite{wilking2011,jana2020nonlinear}. By considering the bacterial biofilm as a polymer-colloid viscoelastic medium \cite{wilking2011,boudarel2018towards,samuel2019regulating,jana2020nonlinear, billings2015material, gordon2017biofilms}, $G^{'}$ should increase with the density of bacteria until reaching a solid-like plateau. Our simulation data point out that the probed $N_b$-range falls into the elastic response regime where $G'$ is not affected by $N_b$ (panel a). On the other hand, the loss modulus $G''$ linearly increases  with $N_b$  (panel d), reflecting the high friction imposed on the crosslinked mesh by dragging large bacterial objects while increasing  density. In agreement with a solid-like behavior lead by the rigidity of the crosslinked mesh, $G''$ is always smaller than $G^{'} $ but approaches larger values compatible with the storage modulus for the largest number of bacteria hereby considered. We conclude therefore about simulated biofilm behavior as a viscoelastic material with a relative high structural rigidity dominant over viscous fluidity (i.e., $G' > G'' $). Otherwise stated, our model predicts the rheological behavior of a quite resilient material with a relatively high mechanical compliance.

{\it Polymer density}. To check the effect of an increasing number of crosslinked polymers on the shear response, we set $N_b=150$, $l_p=100$, $N_s=35000$, $CL=2300$ with an amplitude of $A=75$ $\%$ and period of the deformation of $T=200$ (Fig.\ref{fig:Loss_Storage_Modulus}, panel b and e). When increasing the number of polymers keeping a constant bacterial density, we observe the simulated biofilm to engender a solid-like network with an increasing rigidity with increasing mesh density (panel b). This suggests that bacteria can colonize a preexisting polymer mesh and modify its rigidity according to their space needs; the denser template could entail the more bacteria in a stiffer mesh. However, the out-phase stress response increases only slightly with increasing the number of polymers (panel e). Since this increase of $G''$ is quite small compared to the response caused by the presence of bacteria (panel d), we conclude about $G''$ to be almost independent on the polymer concentration 
\textcolor{black}{as corresponding to our essentially rigid DPD-system describe mesoscopicaly unrelaxed chains within the permanent crosslinks considered.} As referred to the predominantly solid-like behavior emerged with increasing bacterial numbers ($N_b$ in panels a and d), the present results also evidence how biofilm rigidity can be enhanced by increasing polymer numbers ($N_p$ in panels b and e). 
\textcolor{black}{These simulations on the compaction effects induced by polymer density essentially reproduce the nonlinear rheological characteristics experimentally observed for bacterial biofilms of single species with a well structured EPS (i.e., macroscopically homogeneous, compact and continuous) \cite{jana2020nonlinear}.}\\

{\it Solvent effects}. Furthermore, mechanical responses were tested for different hydration levels ($CL=2300$, $N_b=184$, $N_p=80$, $l_p=100$). Upon biofilm deformations of high amplitude ($A=75$ \%), and long period ($T=200$), we observe a hydration effect that makes the storage modulus to decrease with the number of solvent particles (panel c), and preserves regulated friction losses (panel f). Our biofilm DPD- model seems able to absorb the water molecules to mechanically soften at near constant viscosity. Such nonlinear softening emerges as far as averaged distances between bacteria and polymers largely increases by solvent dilution under intense shear deformation thus making the network to become mechanically softer \cite{jana2020nonlinear}. Mesh weakening is however not reflected in the loss modulus because DPD only catches the viscosity of the solvent but not collectively mesoscopic relaxations.

\noindent
\subsection{Rheological dependence on the shear frequency: experiment vs. simulation}

\begin{figure}[h!]
\centering
\includegraphics[width=0.475 \textwidth]{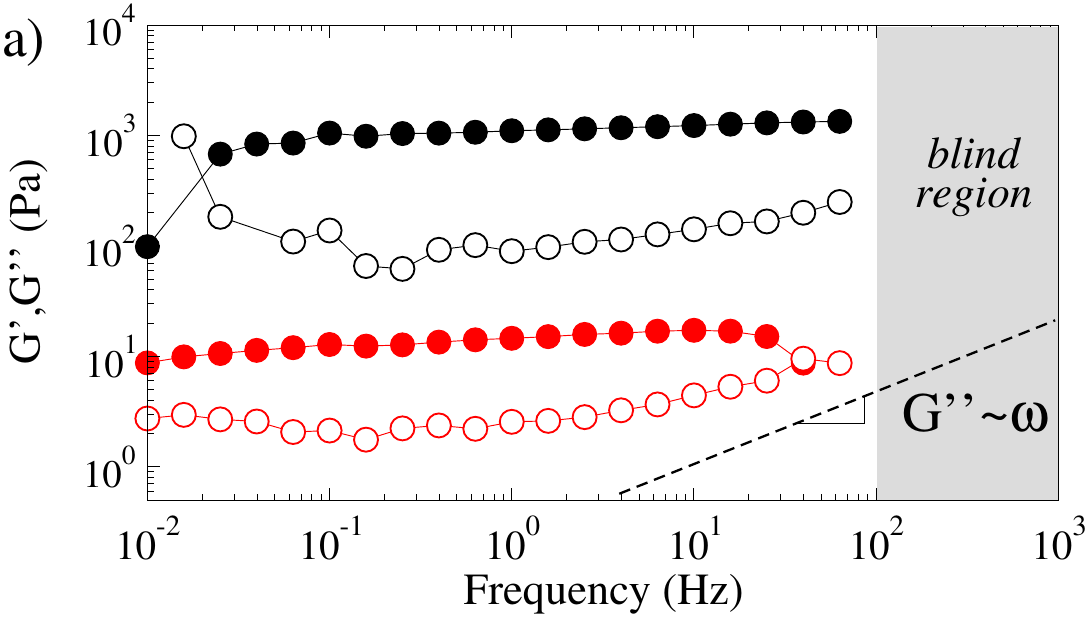}
\includegraphics[width=0.485\textwidth]{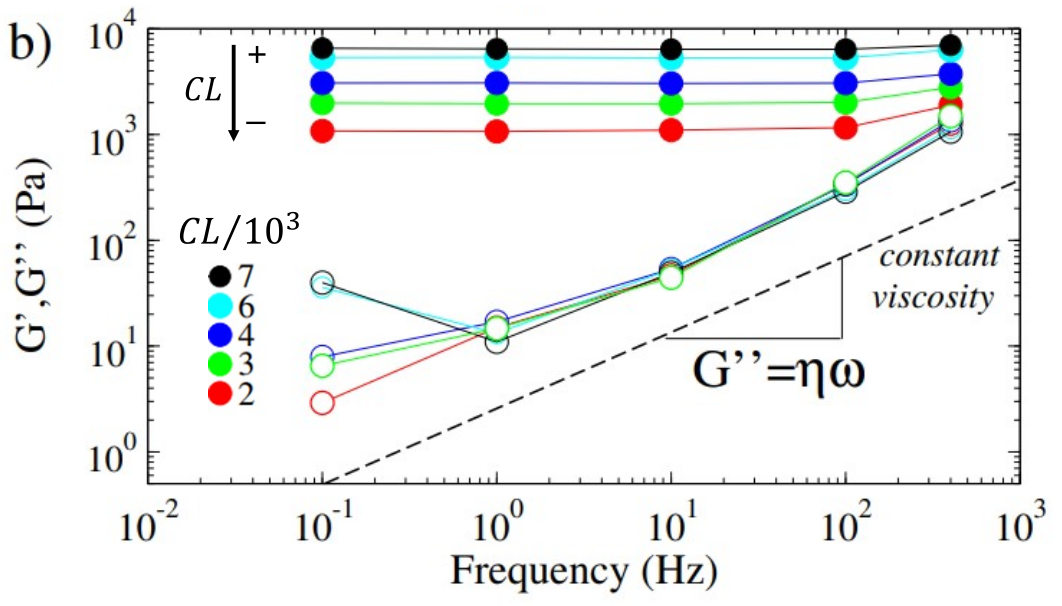}
\caption{a) Experimental measurements of the elastic modulus ($G'$; solid symbols), and loss modulus ($G''$; hollow symbols) for two different classes of {\it P. fluorescens} biofilms: The rigid one grown under shaking conditions (black), and the softer one obtained under static conditions (red); 
\textcolor{black}{measurements performed in triplicate). The symbols represent averaged data replicated in each condition (experimental reproducibility compatible with symbol size)}. The near constant storage modulii evidence soft solid behavior ($G' \sim G_0$). The dashed line indicates the Newtonian trend expected at high frequency for the loss modulus under dominant viscous friction ($G'' \sim \omega$). The grey shadowed blind region is not accessible to experiments (see Methods: Experimental Rheology). b) Synthetic rheology in DPD-biofilms with variable number of permanent crosslinks at the $T_A$-topology: $G'$ (solid symbols) and $G''$ (hollow symbols) computed for the biofilm model. Data refer to $N_b=184$, $N_p=80$, $l=100$, $N_s=35000$, and variable number of crosslinks; from CL = 7000 (upper dataset), down to CL = 2000 (lower dataset); see legend. Whereas the effective rigidities for elastic storage decrease with the number of crosslinks (as observed in experiments; Panel a), ell the calculated frictional losses fall within a common master curve compatible with the constant viscosity encoded in the DPD-simulations (dashed line)\label{fig:fig_exp}
}
\end{figure}

We performed rheological experiments on \textit{P. fluorescens} biofilms grown upon both static and shaken culture conditions (see Experimental Rheology). The viscoelastic responses were subject to dynamic scrutiny under oscillatory shear deformation fixed in the linear regime ($A=1$ $\%$), and variable frequency in the experimentally available window from $\nu= $ 0.01 Hz up to ca. 100 Hz ($\nu=\omega/2\pi$). Figure \ref{fig:fig_exp} compares experimental results (upper panel), and DPD-simulations as performed at equivalent time units (lower panel). The \textit{P. fluorescens} biofilms have both been grown for 24 hours: the first one under static culture conditions (Fig. 7a; red symbols), and the other one under shaking (Fig. 7a; black symbols). Immediately after culture, we place the grown biofilm {\it ex vivo} in the rheometer, and then measure both the linear response $G'$ (solid symbols) and $G''$ (hollow symbols), either for the static biofilm (red symbols), and for the shaking biofilm (black symbols). Due to the technical limitations of the used rheometer, we underline that we could not measure the stress moduli of the \textit{P. fluorescens} biofilm grown under shaking conditions for frequencies beyond the 100 Hz established as a practical ceiling value for experimental macro-rheology. {\color{black} The recorded data intentionally correspond to the linear regime of viscoelastic response ($A = 1\%$), in which a quasi-static response is expected at compatibility with the permanent nature of the rigid (passive) crosslinks considered in the simulations. Further active effects present in the real biofilms as nonlinear softening stresses ($A \geq 10\%$)\cite{jara2021}, re-configurable crosslinks or propulsion bacterium impulses e.g., are not here discussed as not being explictely considered in the current DPD-simulations only capturing quasi-static interactions between "passive" biofilm components in a fixed meshwork of permanent crosslinks}.

The \textit{P. fluorescens} biofilms grown for 24 hours under shaking conditions present higher shear modulii (both $G^{'} $ and $G^{''}$; symbols in black) than those of a biofilm grown under static conditions (symbols in red), independently on the shear frequency probed in the linear regime of rheological response. Otherwise stated, the shaken biofilms develop higher rigidness than those grown under static conditions. Such Hookean stiffening is compatible with the nonlinear strengthening observed under stress (see Fig. 4). Culture shaking might constitute indeed a highly stressed biofilm growth setting which elicits higher crosslinking than static growth, resulting thus into structural stiffening (so much as nonlinear stress results into effective dynamic strengthening). In general, the measured values of the storage modulus are found higher than the frictional losses (especially for low and intermediate frequencies when $G'$ > $G''$). This is the typical behavior of soft solids displaying higher elastic resistance than frictional opposition to shear flow \cite{fiorini2016nanocomposite}. The observed rheological parsimony as a soft solid is characteristic for any \textit{P. fluorescens} biofilm grown longer than 24h (data not shown). For the biofilms grown under static conditions, the rheological experiments show however an increase in frictional losses at the highest frequencies tested  i.e., $G'' \sim \omega$ (see Fig. 7a; dashed line). This dissipative contribution leads a viscoelastic inversion $G'\sim G''$, which is also characteristic for soft solids that strained at high rates make the viscous flow to emerge as a consequence of stress\cite{fiorini2016nanocomposite}. A similar behaviour seems to be extrapolated in the shaken biofilm as far as the difference between $G'$ and $G''$ becomes reduced upon increasing frictional losses at high frequencies whereas the elastic storage remains constant (see Fig. 7a).

To test the capability of our synthetic biofilm model to capture experimental behavior in \textit{P. fluorescens} biofilms, we perform numerical rheology as reported in Fig.\ref{fig:fig_exp}b. After having explored the parameter space in the previous section, we choose to prepare a biofilm with the following parameters: $N_b=184$, $N_p=80$, $l_p=100$, $N_s=35000$ and $A=75$ $\%$, with a variable number of crosslinks spanning the broad range $CL=2000 - 7000$ (considered representative for poor and rich crosslinking, respectively leading soft and stiff elasticity). As previously justified in Ref.\cite{jara2021}, we assume indeed that the main difference between the biofilm grown under static and shaking conditions is only in the matrix composition, being the one of the shaken-grown biofilm the richer in crosslinks.

Figure \ref{fig:fig_exp}.b plots the different datasets computed for $G'$ (closed symbols) and $G''$ (hollow symbols) as a function of the frequency of the shear strain imposed to the simulation box for the different number of crosslinks in the chosen canonical $T_A-$ topology. Accordingly to the established experimental conditions, we numerically compute the shear moduli of the DPD-model by changing the stress frequency over more than four orders of magnitude. Internal simulation units are converted into physical units accordingly to the conversion factors previously described. Consistently with what we observed in experiments, $G'$ softens with decreasing the number of crosslinks but does not show a significant change upon increasing $\omega$, as corresponding to a soft solid with permanent crosslinks unable to undergo relaxation \cite{aufderhorst2019stiffening}. Moreover, the computed values for the storage modulus (closed symbols) are systematically higher than the ones found for the loss modulus (hollow symbols), as expected for soft solids \cite{findley2013creep}. 
\textcolor{black}{The numerical results in Fig. 7.b refer to the $T_A$ topology corresponding to the more open realization facilitating for bacterial reorganization. As previously shown in Fig. 2, these DPD-biofilms are spatially homogeneous in the microscopic terms revealed by the PDF structure. However, not only their rheological properties show to give consistent results when tested varying the parameter space (Fig. 6), but are also capable to qualitatively reproduce the high-frequency (small-scale) rheological behaviour observed in experiments (Fig. 7.a), even if they do not consider spatial inhomogeneity characteristic of real biofilms. Further work is in progress to explicitly mimic spatial inhomogeneities in the numerical DPD-based approach.}

\textcolor{black}{As expected, the simulated frictional losses markedly follow the Dissipative Particle Dynamics as imprinted by the constant solvent viscosity in the whole frequency domain (i.e., $G'' = \omega \eta$; see Fig. 7b). This marked unitary $G'' \approx \omega $ fingerprint actually corresponds to the intrinsic dissipation of the DPD-simulation method which no longer operate as a collective relaxation in the largest mesoscopic scales (at low frequencies).} This result 
\textcolor{black}{is not only compatible with the Newtonian fluidity inherent to the microscopic-DPD} but also agrees with what experimentally observe at high frequencies in both the stimulated shaken biofilms and the statically grown ones (
\textcolor{black}{although they do preserve the high frictional relaxation at low frequencies}; see Fig. \ref{fig:fig_exp}a). However, the DPD-simulated values of $G''$ obtained at the lowest simulation frequencies appear more than two order of magnitudes smaller than $G'$, which indicates a predominantly conservative mechanics under the $\it{quasi}$-static deformation conditions considered in the simulations. The different simulated datasets fall indeed within a same master curve as corresponding to a common viscous friction imparted by the solvent on the single structural entities (independently on the number of crosslinks). The difference is progressively reduced upon increasing $\omega$ (increasing viscous friction under constant rigidity), up to the Newtonian flow behavior $G'\sim G''=\eta\omega$ undergone by the synthetic DPD-biofilms at the higher simulated frequency (corresponding to the faster strain rates). A qualitatively similar behavior occurs in the experiments for \textit{P. fluorescens} biofilms grown under static conditions albeit the difference between $G'$ and $G''$ at low frequencies is not actually so marked as predicted by the simulations. These compared results validate, at least qualitatively, a synthetic capacity of our model DPD-biofilm for predictive rheology in terms of permanent matrix crosslinking. Noticeably, no viscoelastic relaxation can be predicted as far the crosslinks remain fixed across the mapped scales.

\section{Discussion}

Biofilms are composed of bacteria that secrete a mesh of extracellular polymeric substances (EPS) producing a viscoelastic matrix to which they eventually crosslink {\color{black} through of complex (conservative and dissipative) mechanisms that result in a complex rheological response \cite{jara2021}, \cite{boudarel2018towards}, \cite{samuel2019regulating,jana2020nonlinear,gordon2017biofilms,billings2015material}.} Even though a mathematical modelling  of bacterial biofilms might provide a useful tool for controlling biofilm formation, up-to-date modelling cannot be considered completely satisfactory. Tailoring a detailed model that includes physical parameters to mimic hydrodynamics, solute mass transport and active dynamics of the bacterial population within the biofilm is indeed a challenging task. The parametrization of the individual components of a biofilm model increases the complexity of the algorithm beyond limited computational capacity \cite{calude2017deluge}. Therefore, it would be highly desirable to design a structurally simplified model while retaining the relevant dynamics. {\color{black} Further including dynamical details arising from systemic memory effects could be also desirable although computationally unavoidable in the mesoscopic level of complexity requested to capture biofilm rheology as measurable in experiments.}

By taking advantage of a Dissipative Particle Dynamics (DPD) simulation algorithm, we have validated a coarse-grained model of biofilm behavior to numerically explore the rheological properties of a \textit{P. fluorescens} biofilm. The synthetic model is based on the DPD-approach depicted in \cite{jara2021}, and consists of topologically tunable simulation platform of {\color{black} permanent polymer chain crosslinking able to mimic a real bacterial biofilm approaching an effectively passive living component embedded with a variable contents of EPS}. Because DPD models allow to simulate short range interactions of any statically embedded particle, our model enables for progressively increasing structural details at smaller spatial resolution in the polymerized (EPS) mesh; from the macroscopic scale of the rheological response, through the structural mesoscopic details, down to the fluctuating behavior of the single particles under detailed balance of thermal motion against frictional dissipation. This is an important advantage over generally used long-range interaction models\cite{groot2004applications}, which are spatially limited by a continuous scale cut-off that impedes going deeper into smaller levels of detail including the topology of crosslinking and the viscous friction in the underlying microscopic motions. Moreover, the "soft" nature of DPD-interactions allows to increase the integration time by orders of magnitude, allowing to explore rheological time scales normally inaccessible to atomistic simulations. {\color{black} However, our numerical DPD-simulation schema is only weakly dissipative (or too much "conservative") as only considering static permanent crosslinking differently to real biofilms including highly dissipative sinks and sources of mechanical energy e.g., reconfigurable crosslinks, sliding entanglements and propulsive forces within the thriving bacteria. Therefore, our structurally simplified DPD-approach only captures essential features of linear mechanics but fails in describing the active nonlinear response observed in experiments. Whereas the topologically static DPD-simulations naturally exhibit nonlinear strengthening as due to chain ordering under large deformation the real {\it P. fluorescens} biofilms used as a validating setting exhibit contradictory mechanical softening as due to biological activities not considered in the current simulations \cite{jara2021}.} Finally and foremost, our numerical DPD-algorithm explicitly includes hydrodynamics and thermal fluctuations, which are known to play a crucial role at introducing variability in biological systems such as the biofilms here considered \cite{krsmanovic2021hydrodynamics}. {\color{black} Our modelling approach in essence simplify the complexity of real biofilms treating them as composite materials described by a set of physical parameters that recapitulate the principal ingredients of compositional and topological structure. The simulated DPD-biofilms have been constructed as a complex structure of prescribed topology as fixed by the degree and density of crosslinking and known composition by class and number of interacting particles. Because of its mesoscopic nature we have got access to simulate their rheological behaviour as dependent of the treatment processes imposing the history of the shear deformation in a way controlled by the dissipative frictional memory of the microscopic components.}

We have chosen to study biofilms formed by \textit{P. fluorescens} for three main reasons:
1) It is the same model system as explored in our previous study\cite{jara2021}, and other related ones, {\it P. aeruginosa} e.g. (see Ref. \cite{jana2020nonlinear} for a recent review), making it easy to validate our simulation results with already published ones (both experimental and numerical with a more limited scope). 2) The DPD-simulations here validated allow to make predictions in more extended spatio-temporal scales than conventional macro-rheological experiments do allow (including high frequencies and high strain rates often precluded by spurious inertial effects in experiments). 3) Although the model microorganism used in this work for experimental validation of the simulation framework is non-pathogenic, other members of the genus \textit{Pseudomonas} are often pathogenic thus being requested for synthetic analysis {\it in silico}. As a relevant counterexample, \textit{P. aeruginosa} is an opportunistic pathogen that is frequently associated with chronic biofilm infections being hence attractive for simulation forecasting. Overall, the composition of the EPS is similar in both species, with a high proportion of acetylated polysaccharides (alginate-like) and extracellular DNA \citep{Kives, Jennings}, which suggests that the biofilm could be characterised by a similar mechanical behaviour that \textit{P. fluorescens}. The proposed DPD-simulations could make reasonable numerical forecasting on the physical biofilm's fate as performed in "digital tweens" for mechanical behavior, built {\it in silico} as synthetic biofilms resembling the mechanical properties of the real (pathogenic) biofilms.     

\section{Conclusions}
By comparing the experimental to the numerical results, we conclude that our proposed coarse grained model qualitatively reproduces the behavior of rheological moduli over several decades of dynamic behavior. The measured elastic modulus was always higher than the loss modulus both in experiments and simulations as corresponding to soft solid behavior.

Moreover, we predict the decreasing difference between $G'$ and $G''$ observed at higher frequencies as a consequence of dominant viscous friction. This was clearly observed in the softer biofilms prepared under static conditions, and predicted by the DPD-simulations including frictional dissipation as an essential dynamic ingredient. As a very characteristic feature of biofilm rheology, $G'$ and $G''$ tended to converge for high frequency both in numerical and experimental outcomes. Nonetheless, simulations showed a larger dissipative gap between $G'$ and $G''$ than what observed in experiments. In general, when $G' >> G''$ the system flows more like an inviscid liquid whereas the regime where $G'\sim G''$ indicates a viscoelastic system. Hence, our DPD-model presents a marked transition from a rigid solid at low frequency towards a viscoelastic regime at higher frequencies. In the experiments with the real biofilms this transition happens in a less pronounced way more compatible with a soft solid behavior over broader scales than in the simulations.

The qualitative correspondence between real biofilms and synthetic DPD simulations endows the forecasting potential of using a coarse grained DPD approach to biofilm rheology when modelling extremely complex biofilms. Even though the model strongly relies on an {\it a priori} detailed study of the parameter space, in future prospective, this will allow us  to evaluate in more detail the biofilm transitions from a predominantly solid-like system to a predominantly liquid-like behavior, through the interplay of the elastic and loss moduli upon changing the biofilm composition or growth conditions. In fact, this transition can be even more dramatic in biofilms prepared under static conditions and longer maturation times, where the elastic modulus vanishes at high frequency. We are currently working in this direction, studying  
how cross-linking dynamics may play a relevant role in these transitions.

As a prospective outlook, DPD-simulation knowledge on the dynamics of the mechanical transitions within synthetic biofilms may guide future strategies that allow the penetration of antimicrobial agents into “soften” biofilm matrices, even predicting conditions for structural disassembly relevant for technological applications. 

\begin{acknowledgments}

B.O., I.L-M and C.V. acknowledges funding from Grant UCM/Santander PR26/16.
 FM acknowledges funding from Grant PID2019-105606RB-I00, FIS2016-78847-P, PID2019-108391RB-I00, 
and FIS2015-70339 from MINECO; 
REACT-EU program PR38-21-28 ANTICIPA-CM - A grant by Comunidad de Madrid and European Union under FEDER Program; 
from EU in response to COVID-19 pandemics; 
and Comunidad de Madrid under grants S2018/NMT-4389 and Y2018/BIO-5207. 
C.V. from 
PID2019-105343GB-I00 of the MINECO 
F.A  acknowledges support from the ``Juan de la Cierva'' program (FJCI-2017-33580). AKM is recipient of a Sara Borrell fellowship (CD18/00206) financed by the Spanish Ministry of Health. V.B. acknowledges the
support from the European Commission through the Marie
Skłodowska-Curie Fellowship No. 748170 ProFrost. 
The authors acknowledge the computer resources from the Red Española de Supercomputacion (RES) FI-2020-1-0015 and FI-2020-2-0032, and from the Vienna Scientific Cluster (VSC).
\end{acknowledgments}

\section*{Data Availability Statement}
The data that support the findings of this study are available from the corresponding author upon reasonable request.


\section*{References}

%

\end{document}